\documentclass[usenatbib]{mn2e}
\usepackage{graphicx}

\title[Gaussianity Tests of CMB]{Gaussianity of the CMB: Smooth Goodness of Fit
Tests Applied to Interferometric Data}

\author[A. M. Aliaga et al.]{A. M. Aliaga,$^{1,2}$\thanks{e-mail: 
aliaga@ifca.unican.es} 
J. A. Rubi\~no-Mart\'{\i}n,$^3$ 
E. Mart\'{\i}nez-Gonz\'alez,$^1$ 
R. B. Barreiro$^1$ \newauthor
and J .L. Sanz$^1$
\\
$^1$IFCA, CSIC-Univ. de Cantabria, Avda. los Castros, s/n, E-39005 Santander, 
Spain
\\
$^2$Dpto. de F\'{\i}sica Moderna, Univ. de Cantabria, Avda. los Castros, s/n, 
E-39005 Santander, Spain
\\
$^3$Instituto de Astrof\'{\i}sica de Canarias, E-38200 La Laguna, Tenerife, 
Spain
}

\pagerange{\pageref{firstpage}--\pageref{lastpage}} \pubyear{2004}

\begin{document}
\date{}
\label{firstpage}

\maketitle

\begin{abstract}
We adapt the smooth tests of goodness of fit developed by 
\citet{rayner1,rayner2} to the study of the non-Gaussianity of interferometric 
observations of the cosmic microwave background (CMB). The interferometric 
measurements (visibilities) are transformed into signal-to-noise eigenmodes, 
and then the method is applied directly in  Fourier space. This 
transformation allows us to perform the analysis in different subsets of 
eigenmodes according to their signal-to-noise level. The method can also deal 
with non-uniform or incomplete coverage of the $UV$ plane. We explore here two 
possibilities: we analyse either the real and imaginary parts of the 
complex visibilities (Gaussianly distributed under the Gaussianity hypothesis) 
or their phases (uniformly distributed under the Gaussianity hypothesis). 
The power of the method in discriminating between Gaussian and non-Gaussian 
distributions is studied by using several kinds of non-Gaussian simulations. 
On the one hand, we introduce a certain degree of non-Gaussianity directly into
the Fourier space using the Edgeworth expansion, and afterwards the desired 
correlation is introduced. On the other hand, we consider interferometric 
observations of a map with topological defects (cosmic strings). 
To these previous non-Gaussian simulations we add different noise levels 
and quantify the required signal-to-noise ratio necessary to achieve a 
detection of these non-Gaussian features. 
Finally, we have also studied the ability of the method to 
constrain the so-called nonlinear coupling constant $f_{NL}$ using 
$\chi^2$ simulations.
The whole method is illustrated here by application to simulated data 
from the Very Small Array interferometer.

\end{abstract}


\section{Introduction}

The study of the probability distribution of the primordial density 
fluctuations in the Universe is one of the most fundamental challenges in the
present day cosmology. 
The determination of this probability distribution
could constrain the ensemble of cosmological theories of the formation of 
primordial density fluctuations. In particular, standard inflationary theories 
establish that these density fluctuations are Gaussianly 
distributed \citep{guth}. So the detection of some non-Gaussian features 
would question these kinds of theories. 
A powerful tool in the determination of cosmological parameters is the cosmic
microwave background (CMB). The primordial density fluctuations left their 
imprint in the CMB because of the thermal equilibrium existing before  
matter--radiation decoupling. In this manner, the initial density fluctuations 
led to anisotropies in the CMB, and these inherited the probability 
distribution of the former. In short, testing CMB Gaussianity is equivalent 
to testing the Gaussianity of the primordial fluctuations and hence the 
validity of standard inflation.

For these reasons there is great interest in the implementation of
statistical methods to analyse the Gaussianity of current and 
future CMB data. A representative sample of these methods 
(containing real-space statistics, wavelets and Fourier-based analyses)
have been recently applied to probe the Gaussianity of the data from the 
WMAP mission \citep{komatsu,chiang,eriksen04a,eriksen04b,vielva,park,cruz}.

In this paper, we shall concentrate on the smooth tests of goodness of fit 
developed by \citet{rayner1,rayner2}. These methods have been already adapted 
to CMB data analysis, and applied \citep{cayon} to the study of the Gaussianity
of MAXIMA data \citep{balbi, hanany}. We further adapt this method here to 
deal with data from interferometers. 

With their intrinsic stability (only the correlated 
signal is detected) and their ability to reject atmospheric signals (e.g. 
\citealt{int33,lay00}) interferometers have proved their worth in ground-based 
observations of the CMB. Over the last
few years several interferometers have provided
high-sensitivity measurements of the power spectrum at intermediate
angular scales. In particular, these experiments include the Very Small 
Array (VSA, \citealt{watson03}), DASI \citep{dasi} and CBI \citep{cbi};
for a more detailed compilation of interferometric experiments 
measuring the CMB,  e.g. \citet{white}. These instruments sample 
the Fourier modes of the measured intensity, so they provide a direct 
measurement of the power spectrum on the sky and allow us to study the 
Gaussianity directly in harmonic space.

Non-Gaussianity analyses have been already performed on interferometer
datasets \citep{savage,smith}. In the first case these analyses are 
mainly based on real-space statistics (they are applied to the 
maximum-entropy reconstructed maps in real space); while in the second case, 
the analysis is performed in Fourier space by computing the bispectrum in 
a similar way to the computation of the power spectrum. 
We follow the second approach and  present here the smooth tests of 
goodness-of-fit adapted to the study of the Gaussianity of interferometric 
data directly in the visibility space, so we test
directly  the Gaussian nature of the Fourier modes on the sky. 

Since an interferometer measures complex quantities ({visibilities}),
we present two different analyses in this paper, one  on the real and
imaginary parts, and another  on the phases. 
The experimental data are a combination of the signal plus noise. 
In principle, it would be desirable to analyse data in which signal 
dominates over the noise; that is, data with a high
signal-to-noise ratio. One suitable approach to select only data with high 
signal-to-noise ratio is the transformation of the data into signal-to-noise 
eigenmodes \citep{bond}. This formalism allows us to eliminate the data 
dominated by noise. Moreover, as we explain below, the signal-to-noise 
eigenmodes are not correlated, allowing the direct implementation of the
method.

As it is shown below, our method looks for deviations with respect 
to the null hypothesis (Gaussianity) in statistics related to the measured 
moments (mean, variance, skewness, kurtosis, ...) of the visibility 
signal-to-noise eigenmodes. Thus, it is clear that the test is specially 
tailored for detecting non-Gaussian signals directly in the Fourier domain. 
This Gaussianity analysis of the Fourier components completes the analysis 
of Gaussianity in the real space, given that some non-Gaussian detection 
in the Fourier space could indicate some degree of non-Gaussianity 
in the real space and vice versa (a linear transformation of Gaussian 
variables is also Gaussian). In fact, for example, the analysis of Gaussianity 
directly in the Fourier space by means of the bispectrum has demonstrated to be
a powerful tool in the detection of the $f_{NL}$ parameter 
\citep{komatsu2001,komatsu}. Then, the analysis in
Fourier space could detect non-Gaussianity that, in principle, should be
more evident in real space. As an illustration of this fact, we will show how 
our method detects the non-Gaussianity of a cosmic strings map directly in 
Fourier space, even though it seems more natural to search the non-Gaussian 
features of these objects in real space.

The organization of the paper is as follows. Smooth tests of goodness-of-fit
are introduced in Section \ref{section_2}. We focus our interest in the 
statistics developed by \citet{rayner1,rayner2} and their application to the
Gaussian and  uniform distributions that we are going to test. 
In Section \ref{section_3} signal-to-noise eigenmodes are reviewed and 
applied to the case of interferometer observations. The simulations we 
are using are described in Section \ref{sec:vsa}, where the Very Small Array
(VSA) is taken as a reference experiment. The application to 
Gaussian simulations is described in Section \ref{sec:gauss}.
The power of the test in discriminating between 
Gaussian and non-Gaussian signals affected by Gaussian noise is studied in 
Section \ref{section_4}. 
We consider three kinds of non-Gaussian simulations: 
those performed using the Edgeworth expansion introduced 
by \citet{martinez}, a string simulation created by \cite{bouchet} 
and $\chi^2$ simulations with a $f_{NL}$ coupling parameter. 
Finally, Section \ref{section_5} is dedicated to discussion and conclusions.

\section{Smooth tests of goodness-of-fit} \label{section_2}

In this section we present the smooth tests of goodness-of-fit and the work of
\citet{rayner1,rayner2} applied to a Gaussian and a uniform variable.

Let us suppose $n$ independent realizations, $\{x_i\}$, of a statistical 
variable $x$ ($i=1,\ldots,n)$. Our aim is to test whether the probability 
density function of $x$ is equal to a prefixed function $f(x)$. The smooth 
tests of goodness-of-fit are constructed to discriminate between the 
predetermined function $f(x)$ and another  that deviates smoothly from the 
former and is called the \emph{alternative density function}. We 
consider an alternative density function given by $f(x,\bmath{\theta})$, 
where $\bmath{\theta}$ is a parameter vector whose $i$th component 
is $\theta_i$ and there exists a value $\bmath{\theta}_0$ such that 
$f(x,\bmath{\theta}_0)=f(x)$. The alternative function deviates 
smoothly from $f(x)$ when $\bmath{\theta}$ is displaced from 
$\bmath{\theta}_0$.

The probability density function of the $n$ independent realizations of $x$ is 
given by $\prod_{i=1}^{n} f(x_i,\bmath{\theta})$. Given these realizations, we 
calculate the estimated value of $\bmath{\theta}$ by means of the maximum
likelihood method and denote this value by $\bmath{\hat{\theta}}$. The 
quantity $W$ is defined such that

\[
\exp \bigg( \frac{W}{2} \bigg) = \frac{\prod_{i=1}^{n} 
f(x_i,\bmath{\hat{\theta}})}{\prod_{i=1}^{n} f(x_i,\bmath{\theta}_0)} , 
\]

\noindent where $W$ is a measurement of the difference between 
$\bmath{\hat{\theta}}$ and $\bmath{\theta}_0$ and is therefore a test of the
hypothesis that $\bmath{\theta}=\bmath{\theta}_0$. Assuming 
$\bmath{\hat{\theta}}$ to be close to $\bmath{\theta}_0$ and $n$ to 
be very large ($n \rightarrow \infty$), the 
quantity $W$ is equal to the so-called \emph{score statistic} 
\citep{cox,aliaga3b}, 

\begin{equation}\label{eq:001}
S = \bmath{U}^\rmn{t}(\bmath{\theta}_0) \mathbfss{I}^{-1}(\bmath{\theta}_0) 
\bmath{U}(\bmath{\theta}_0),
\end{equation}

\noindent where $\bmath{U}(\bmath{\theta})$ is the vector whose components are 
$U_i(\bmath{\theta}) \equiv \partial \ell(\{ x_j\}, \bmath{\theta}) 
/ \partial \theta_i$, with the log-likelihood function defined as 
$\ell (\{ x_j\},\bmath{\theta}) \equiv 
\log \prod_{i=1}^{n} f(x_i,\bmath{\theta})$. $\bmath{U}^\rmn{t}$ is the 
transposed vector of $\bmath{U}$ and the components of the matrix 
$\textbfss{I}$ are equal to $I_{ij}(\bmath{\theta}) \equiv 
\langle U_i(\bmath{\theta})U_j(\bmath{\theta}) \rangle$. 

Among all the possible choices of alternative density functions, we select that
 presented in the work of \citet{rayner1,rayner2}.

\subsection{Rayner--Best Test}

In the work of \citet{rayner1,rayner2}   an alternative density 
function is defined  as

\[
g_k(x)= C(\theta_1,\ldots,\theta_k) \exp \bigg\{ \sum_{i=1}^k \theta_i h_i(x) 
\bigg\} f(x),
\]

\noindent where $C(\theta_1,\ldots,\theta_k)$ is a normalization constant and 
the $h_i$ functions are orthonormal on $f$ with $h_0(x)=1$ (note that 
$\bmath{\theta}_0=0$). It can be demonstrated that the score statistic 
associated with the $k$ alternative is given by

\begin{equation}\label{eq:004}
S_k = \sum_{i=1}^k U_i^2, \qquad \textrm{with} \qquad U_i=\frac{1}{\sqrt n} 
\sum_{j=1}^n h_i(x_j).
\end{equation}

Note that the previous $U_i$ quantity is not the  $i$th component of the 
vector $\bmath{U}$ in expression (\ref{eq:001}) (although they are 
closely related). We denote the two quantities with the same letter to 
follow the notation of \citet{rayner1,rayner2}. 
As equation (\ref{eq:001}) shows, the expression (\ref{eq:004}) gives 
the score statistic only when the distribution of the data is $f(x)$ 
(because the condition $\bmath{\theta}= \bmath{\theta}_0$ is imposed). In this 
case (that is, when the distribution of the data is equal to the prefixed one),
the $S_k$ statistic is distributed as $\chi_k^2$ when $n \rightarrow \infty$.
This holds because $U_i$ is Gaussianly distributed when $n \rightarrow \infty$ 
(the central limit theorem). Moreover, it is easy to prove that the mean value 
of $U_i^2$ is equal to unity, independently of the value of $n$. We see that 
we can work with the $S_k$ quantities or alternately with  $U_i^2$  (if 
the distribution of the data is equal to $f(x)$, these are distributed 
following a $\chi_1^2$ function). 

In this paper we apply these statistics to two cases: on one hand,  
to the real and imaginary parts of the visibilities and on the other hand
to the phases of the visibilities. In the first case, the function $f(x)$ is a 
Gaussian distribution, and in the second case the distribution is a uniform.

\subsection{Gaussian Variable}\label{sec:001}

If $f(x)$ is a Gaussian distribution with zero mean and unit variance, the 
$h_i$ functions are the normalized Hermite--Chebyshev polynomials 
$h_i(x)=P_i(x)/s_i$ with $s_i=\sqrt{i!}$, $P_0(x)=1$, $P_1(x)=x$, and, 
for $i \ge 1$, $P_{i+1}(x)=xP_i(x)-iP_{i-1}(x)$. The $U_i^2$ statistics are
given by:
\begin{eqnarray}
U_1^2  &=& n ( \hat{\mu}_1)^2                                     \nonumber\\
U_2^2  &=& n ( \hat{\mu}_2 - 1 )^2 /2                             \nonumber\\
U_3^2  &=& n ( \hat{\mu}_3 - 3 \hat{\mu}_1 )^2 /6                 \nonumber\\
U_4^2  &=& n [(\hat{\mu}_4 - 3 ) - 6 (\hat{\mu}_2 -1) ]^2  /24     \nonumber
\end{eqnarray}

\noindent where $\hat{\mu}_{\alpha}=(\sum_{j=1}^n x_j^{\alpha})/n$ is the 
estimated moment of order $\alpha$. Bearing in mind the relation between 
the $U_i^2$ and the $S_k$ quantities, we see that the statistic $S_k$ is 
related to moments of order $\le k$, so that this test is \emph{directional}; 
that is, it indicates how the actual distribution deviates from Gaussianity. 
For example, if $S_1$ and $S_2$ are small and $S_3$ is large. The data then
have a large $\hat{\mu}_3$ value because of the relation between $\hat{\mu}_3$ 
and $S_3$. Then, the data have a large skewness value. This  holds for all
distributions and not only for the Gaussian case.

These expressions (or the equivalent $S_k$) are used in \citet{cayon} to 
test the Gaussianity of the MAXIMA data. The skewness and the kurtosis of 
these data are constrained in \citet{aliaga3a} using the same method.

\subsection{Uniform Variable} \label{sec:unif}

The smooth test of goodness-of-fit applied to the uniform distribution was 
developed by the first time by \citet{neyman}, and his work is the starting 
point of the work of \citet{rayner1,rayner2}. If we want to test whether 
$f(x)$ is a uniform distribution in the interval $[0,1]$, we define 
$z=2x-1$ and then we test if $z$ 
is a uniform variable in $[-1,1]$. In this case we take the Legendre 
polynomials: $P_0(z)=1$, $P_1(z)=z$ and 
$(i+1) P_{i+1}(z)=(2i+1)zP_{i}(z)-i P_{i-1}(z)$ for $i \ge 1$. Thus the 
normalized functions are $h_i(z)=P_i(z)/s_i$ with $s_i=(2i+1)^{-1/2}$. In 
this case:

\begin{eqnarray}
U_1^2  &=&  3 n (\hat{\mu}_1)^2                                    \nonumber\\
U_2^2  &=&  45 n (\hat{\mu}_2 - 1/3 )^2 /4                        \nonumber\\
U_3^2  &=&  7 n (5\hat{\mu}_3 - 3 \hat{\mu}_1 )^2 /4              \nonumber\\
U_4^2  &=&  9 n [ 35(\hat{\mu}_4 - 1/5 ) - 30 (\hat{\mu}_2 -1/3) ]^2 /64
                                                       \nonumber
\end{eqnarray}

\noindent where $\hat{\mu}_{\alpha}=(\sum_{j=1}^n z_j^{\alpha})/n$. As in the
Gaussian variable case, we see that the test is directional: deviations in the 
estimated moments $\hat{\mu}_{\alpha}$ from the expected values for a uniform 
distribution give deviations in the statistics.

\section{Smooth Goodness-of-Fit Tests and Signal-to-noise eigenmodes:
Application to interferometer observations} \label{section_3}

If we assume a small field size (where the flat-sky
approximation of the observed region is valid), an interferometer measures 
complex visibilities at a frequency $\nu$ (the van Cittert--Zernicke theorem):
\begin{equation} \label{eq:005}
V(\bmath u, \nu)=\int P(\bmath x, \nu) B(\bmath x, \nu) 
\exp{(i 2 \pi \bmath u \cdot \bmath x)} \rmn d^2 \bmath x,
\end{equation}

\noindent where $\bmath x$ is the angular position of the observed point on 
the sky and $\bmath u$ is a baseline vector in units of the wavelength of the
observed radiation ( monochromatic radiation is assumed, and hereinafter we
 omit the dependence on the frequency, $\nu$). $P(\bmath x)$ is 
the primary beam of the antennas (normalized to $P(0)=1$), and the function 
$B(\bmath x)$ is the brightness distribution on the sky, which 
can be easily translated into temperature contrast values 
$\Delta T(\bmath x) /T$ for the case of the CMB.

If we denote by $\widetilde{P}$ the Fourier transform of $P$ and 
by $C(u)$ the CMB power spectrum (in units of brightness, $u=|\bmath u|$) then 
the correlations between the visibilities observed at $\bmath u_i$ and 
$\bmath u_j$ are given by 

\begin{eqnarray}\label{eq:002}
\langle V(\bmath u_i) V^*(\bmath u_j) \rangle &=& \int \widetilde{P} 
(\bmath u_i - \bmath u)\widetilde{P}^* (\bmath u_j - \bmath u) C(u) 
\rmn d^2 \bmath u     \nonumber \\
\langle V(\bmath u_i) V(\bmath u_j) \rangle &=& \int \widetilde{P} 
(\bmath u_i - \bmath u)\widetilde{P} (\bmath u_j + \bmath u) C(u) 
\rmn d^2 \bmath u
\end{eqnarray}
\noindent where $\widetilde{P}^*$ denotes the complex conjugate of 
$\widetilde{P}$.

These quantities can be computed semi-analytically as described in
\citet{hobson}, who show how to perform a maximum-likelihood
analysis to extract the CMB power spectrum from interferometer observations. 
Note that for the common case in which the primary beam can be approximated 
by a Gaussian function $P(\bmath x)=\exp (- |\bmath x|^2/2\sigma^2)$, it 
can be demonstrated that the real and the imaginary parts of the observed 
complex visibilities are uncorrelated and thus 
independent, if the Gaussianity of the CMB holds.

Our aim is to test if the visibilities are Gaussianly distributed. This can be 
done by analysing the real and imaginary parts of the visibilities or studying
their phases. The following subsections explain how we work with the data.

\subsection{Signal-to-Noise Eigenmodes}\label{sec:002}

We work with signal-to-noise eigenmodes as they are defined in the work of 
\citet{bond}. Let us suppose an observed variable $\Delta_k$, where the 
subscript $k$ denotes a pixel or a position in the space where that variable is
defined (for example the real or the imaginary parts of the visibility 
$V(\bmath u_k)$ are measured in the position $\bmath u_k$ of the called $UV$ 
space). This variable is the sum of a signal and a noise component 
($\Delta_k = s_k+n_k$)
whose respective correlation matrices have components given by 
$C_{s,kk'} = \langle s_k s_{k'}\rangle$ and 
$C_{n,kk'} = \langle n_k n_{k'}\rangle$. The brackets indicate average to many 
realizations. The signal-to-noise eigenmodes are defined by

\begin{equation} \label{eq:003}
\xi_k = \sum_p (\mathbfss{R} \mathbfss{L}_n^{-1})_{kp} \Delta_p,
\end{equation}

\noindent where $\mathbfss{L}_n$ is the called \emph{square root matrix} of the
noise correlation matrix (i.e. $\mathbfss{C}_n = \mathbfss{L}_n 
\mathbfss{L}_n^\rmn{t}$) and  $\mathbfss{R}$ is the rotation matrix which 
diagonalizes the matrix 
$\mathbfss{L}_n^{-1} \mathbfss{C}_s \mathbfss{L}_n^{-\rmn{t}}$. 
The eigenvalues of this diagonalization are denoted by $E_k$. In the case we 
are studying, the noise has zero mean and is not correlated, i.e.\ 
$\langle n_k n_{k'}\rangle = \sigma_k^2 \delta_{kk'}$, so the components 
of $\mathbfss{L}_n$ are $L_{n,kk'} = \sigma_k \delta_{kk'}$.

Eq. (\ref{eq:003}) gives us \emph{transformed} 
signal and noise $\tilde{s}_k$ and $\tilde{n}_k$ such that: 
$\xi_k = \tilde{s}_k + \tilde{n}_k$ with 
$\langle \tilde{s}_k \tilde{s}_{k'}\rangle = E_k\delta_{kk'}$ and 
$\langle \tilde{n}_k \tilde{n}_{k'}\rangle = \delta_{kk'}$. So, the correlation
matrix of $\xi_k$ is given by $\mathbfss{C}_{\xi} = 
\mathbfss{L}_{\xi} \mathbfss{L}_{\xi}^\rmn{t}$ with $L_{\xi,kk'}= (E_k+1)^{1/2}
\delta_{kk'}$. We now have a clear characterization of the  
signal-to-noise relation of our (transformed) data. The noise dominates over 
the signal if $E_k < 1$. So signal-to-noise eigenmodes $\xi_k$ with a very 
low value of the associated eigenvalue $E_k$ are very much dominated 
by the noise and we do not want to include them in our analysis. This 
is the main interest of this approach, but there is another interesting 
point: the signal-to-noise eigenmodes are uncorrelated data. This latter 
fact will be very useful in the application of the tests of Gaussianity 
described in this paper as explained in the next subsection.

\subsection{Real and Imaginary Parts}\label{sec:004}

We write the visibilities in terms of their real and imaginary parts:
\[
V(\bmath u_i)=V_i^R + iV_i^I
\]

Testing the Gaussianity of the visibilities is equivalent to testing the
joint Gaussianity of their real and imaginary parts. As indicated above,
when the primary beam is Gaussian, the real and imaginary parts are 
independent if the Gaussianity of the CMB holds.

Our data to be analysed are the set of real parts of the visibilities which are
correlated among them (that is: $\langle V_i^R V_j^R \rangle \neq 0$) and 
imaginary parts which also are correlated 
($\langle V_i^I V_j^I \rangle \neq 0$). Moreover, we only have one realization.
The test here presented works with a large amount of independent data (see the 
hypothesis in the deduction of the score statistic, section 2). To convert our 
correlated (and then dependent) data to a sample of independent data we proceed
in the following way. Given the real part of the visibilities, we perform the 
transformation in signal-to-noise eigenmodes (expression \ref{eq:003}) and 
define the variable 
\begin{equation}\label{eq:y_k}
y_k = \sum_j L^{-1}_{\xi,kj} \xi_j =\xi_k/(E_k+1)^{1/2}. 
\end{equation}

These new data are uncorrelated and normalized (zero mean, unit dispersion). We
operate in the same way with the imaginary parts and add the resulting 
quantities to those obtained with the real parts. Finally, we have the $y_i$
data ($i =1,\ldots,2\times$number of visibilities) which are uncorrelated and 
normalized. Moreover, if the visibility distribution is multinormal then 
the ${y_i}$ data are independent and Gaussian distributed with zero mean and 
unit dispersion. Then, if Gaussianity holds, the $y_i$ data fulfill the
hypothesis of the smooth tests of goodness-of-fit and their $U_i^2$ statistics 
must have a $\chi^2_1$ distribution. The test is then applied to the ${y_i}$ 
quantities.

\subsection{Phases} \label{sec:3_3}

The analysis can be performed on the visibility phases. 
It is well-known that if the visibility, $V(\bmath u_i)$, is Gaussianly 
distributed, its phase has a uniform distribution. 
The phases of the visibilities are not independent because there 
are correlations of the real parts among themselves, and among the
imaginary parts (the real and imaginary parts are not correlated if 
the primary beam is a Gaussian function).
Thus, and for the case of a Gaussian primary beam, 
given two points $\bmath u_1$ and $\bmath u_2$ in the $UV$ plane, the
joint probability of finding the phase value $\theta_1$ 
in $\bmath u_1$ and 
$\theta_2$ in $\bmath u_2$ is given by

\[
P(\theta_1,\theta_2)= E^2 F \left( \frac{1}{\alpha \beta} + 
\frac{\gamma k_3}{(\alpha \beta - \gamma^2)^{3/2} } \right), 
\]

\noindent where:

\begin{eqnarray}
E      &=&  (AC-B^2)(AC-D^2)                                      \nonumber\\  
F      &=&  (2\pi)^{-2} E^{-1/2}                                 \nonumber\\
A      &=&  \langle (V_1^R)^2 \rangle = \langle (V_1^I)^2 \rangle  \nonumber\\
B      &=&  \langle V_1^R V_2^R \rangle                            \nonumber\\
C      &=&  \langle (V_2^R)^2 \rangle = \langle (V_2^I)^2  \rangle  \nonumber\\
D      &=&  \langle V_1^I V_2^I \rangle          \nonumber\\
\alpha &=&  C(AC-D^2 \cos^2 \theta_1 - B^2 \sin^2 \theta_1)    \nonumber\\
\beta  &=&  A(AC-D^2 \cos^2 \theta_2 - B^2 \sin^2 \theta_2)      \nonumber\\
\gamma &=&  B(AC-D^2) \cos\theta_1 \cos\theta_2 +   \nonumber\\
       & &  D(AC-B^2)\sin \theta_1 \sin \theta_2      \nonumber\\
k_3    &=& \frac{\pi}{2}+ \arctan \frac{\gamma}{(\alpha \beta -\gamma^2)^{1/2}}
 +\frac{(\alpha \beta -\gamma^2)^{1/2}\gamma}{\alpha \beta }    \nonumber
\end{eqnarray}

\begin{figure}
\begin{center}
\includegraphics[width=85mm]{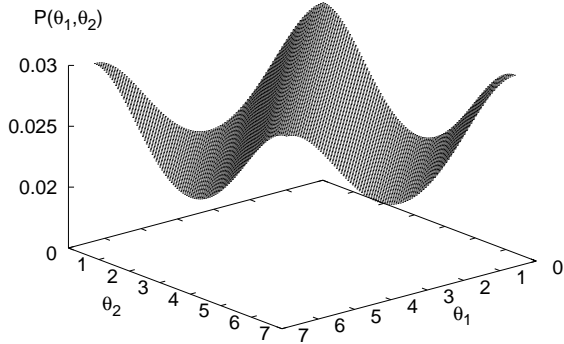}
\caption{Joint probability of finding the values of the phases $\theta_1$ and 
$\theta_2$ in the positions $\bmath u_1$ and $\bmath u_2$.}
\label{fig1}
\end{center}
\end{figure}

Thus $P(\theta_1,\theta_2) \ne P(\theta_1) \cdot  P(\theta_2) = (1/2\pi)^2
\approx 0.0253$. As an example, we have calculated the form of the previous 
probability for the particular case $A=C=1$ and $B=D=0.1$. The plot is shown
in Figure \ref{fig1}.

The statistics of subsection 2.3 cannot be applied directly to the phases
because these are not independent. We therefore construct a related 
independent quantities in the following way. We decorrelate the different 
visibilities so that, under the assumption of Gaussianity, they will be 
independent. To do this we proceed as in the previous section: 
we decorrelate the real parts of the 
visibilities. That transformation gives us the quantities $y^R_i$. We proceed 
in the same way with the imaginary parts and  obtain the associated 
quantities  $y^I_i$. In this way, we have the complex number $y^R_i + i y^I_i$.
These quantities are independent for different $i$. We calculate the phase of 
each one and apply to them the statistics explained in subsection 2.3 for a 
uniform variable.

\section{Simulated interferometer observations} \label{sec:vsa}

\begin{figure}
\begin{center}
\includegraphics[width=\columnwidth]{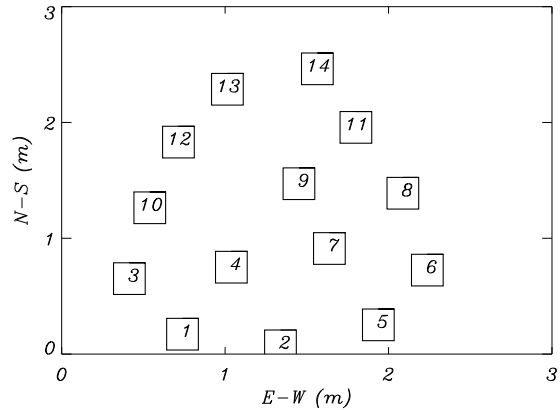}
\caption{The VSA extended array since October 2001 in its extended
configuration. There are fourteen antennas mounted at an angle of 
35$^\circ$ to a tilting table hinged along its northern edge.}
\label{fig_ant}
\end{center}
\end{figure}

To illustrate the method described above, in what follows we apply it to
simulated observations with the Very Small Array. 
The VSA is a 14-element heterodyne interferometer array, 
operating at frequencies between 28 and 36 GHz with a 1.5 GHz bandwidth 
and a system temperature of approximately 30~K, sited at 2400~m 
altitude at the Teide Observatory in Tenerife (see \citealt{watson03} for a
detailed description).
It can be operated in two complementary configurations, referred as
the compact array (which probes the angular range
$\ell \approx$ 150--900; see \citealt{taylor03} for observations 
in this configuration) and the extended array ($\ell$ = 300--1500; 
see \citealt{grainge03} and \citealt{dickinson} for observations in this
configuration).
For the simulations in this paper, we will use as a template experiment
the extended array configuration, which corresponds to the
antenna arrangement shown in Figure \ref{fig_ant}.
The observing frequency is 33~GHz, and the primary beam at this
frequency has a FWHM of 2.1$^\circ$ (corresponding to a $1/e$ diameter 
of the aperture function of $\Delta u \sim 24\lambda$). 
For the VSA, the shape of the antenna can be approximated by a Gaussian 
function, so we can calculate the correlation matrix using the 
expressions (11) and (12) presented in \citet{hobson}.

We use a template observation corresponding to $N_{\rm days}\times$4~hr 
integration time, and we explore several values of $N_{\rm days}$.  
The corresponding noise level will be simulated by adding to each 
visibility a random Gaussian number with an amplitude of $\sim$8~Jy per 
visibility (one channel, one polarization) in an integration time of 
1~min, to reproduce the observed sensitivity of the VSA. Thus, in a 
single day's 4~hr observation we get an rms of $\sim$54~mJy/beam in 
the real-space map. A simulated 1 day VSA file typically contains ~25000 
64 s visibilities. In order to perform the analyses here, we proceed to 
bin these data into cells of a certain size, as described in \cite{hobson}. 
For the analyses carried out in this paper, we have used a bin size 
of 9$\lambda$, where $\lambda$ is the wavelength of the measured signal, 
which is similar to that used for the power spectrum evaluation in 
\cite{grainge03}. We have explored different values of the bin size, and the 
number has been chosen according to two criteria. If the bin size is too 
small, the signal-to-noise ratio per pixel is very low.
On the other hand, for large bin sizes,
the number of binned visibilities is too small to apply the test.
The value of 9$\lambda$ provides both a reasonable number of
visibilities and good power in the detection. Nevertheless, we have
checked that small variations in this number (using values from 9 to 
14$\lambda$) do not produce significant changes.

Our template observation of the VSA contains a total of 895 visibility points 
after binning in 9$\lambda$ cells. This template will be used throughout the
paper.

\section{Gaussian simulations}
\label{sec:gauss}

In this section we calibrate the method by using simulated Gaussian
observations. 
It can be shown that the correlation matrix of the real parts of the 
visibilities can be written as $\mathbfss{C}_s^R=\mathbfss{L}_s^R 
(\mathbfss{L}_s^R)^t$ with $L_{s,ij}^R = \sigma_i (E_j^R)^{1/2} R^R_{ji}$, 
where $E_j^R$ are the signal-to-noise eigenvalues and $\mathbfss{R}^R$ is the
rotation matrix defined in subsection \ref{sec:002}. To generate Gaussian 
simulations with the desired correlation we start with a set $y_i \sim N(0,1)$ 
($i =1,\ldots,$number of visibilities). The real parts of the visibilities are 
given by $V_i^R = \sum_j L_{s,ij}^R y_j$. In an analogous way we construct 
$V_i^I$. After that, a Gaussian realization of the noise is added to each 
visibility.

\begin{figure*}
\begin{center}
\includegraphics[width=8cm]{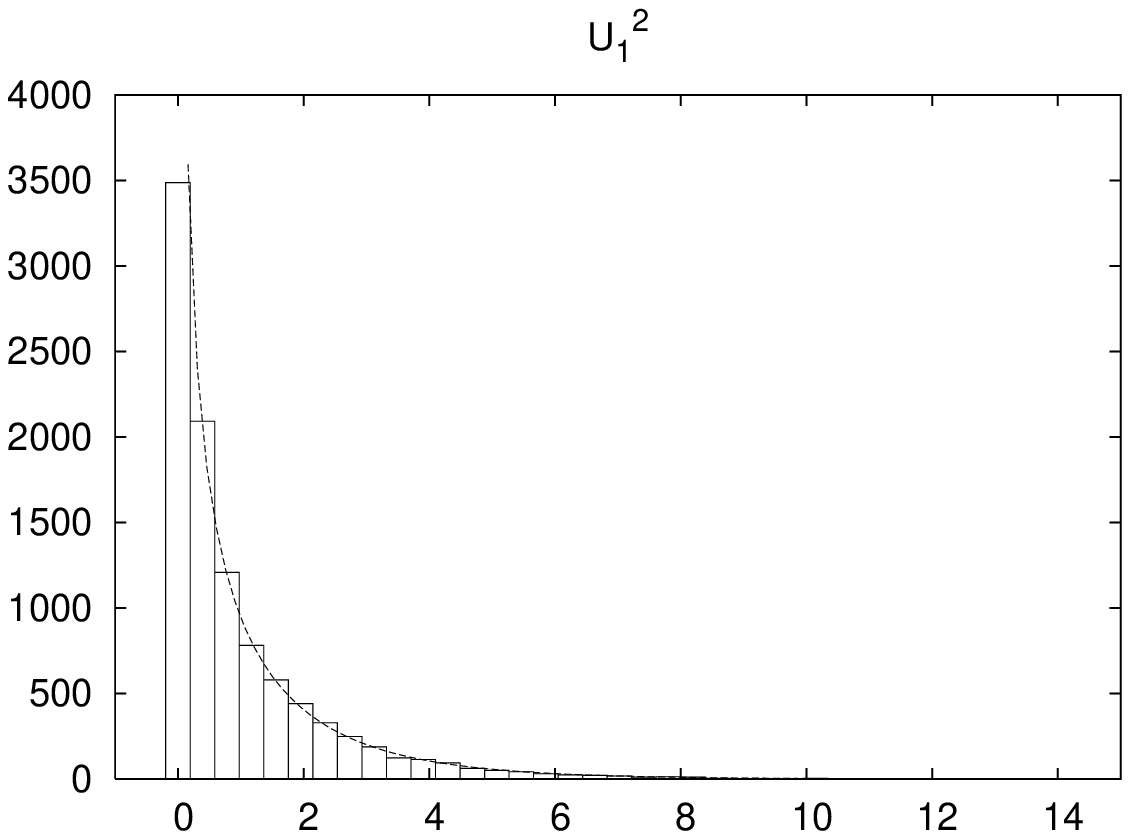}
\includegraphics[width=8cm]{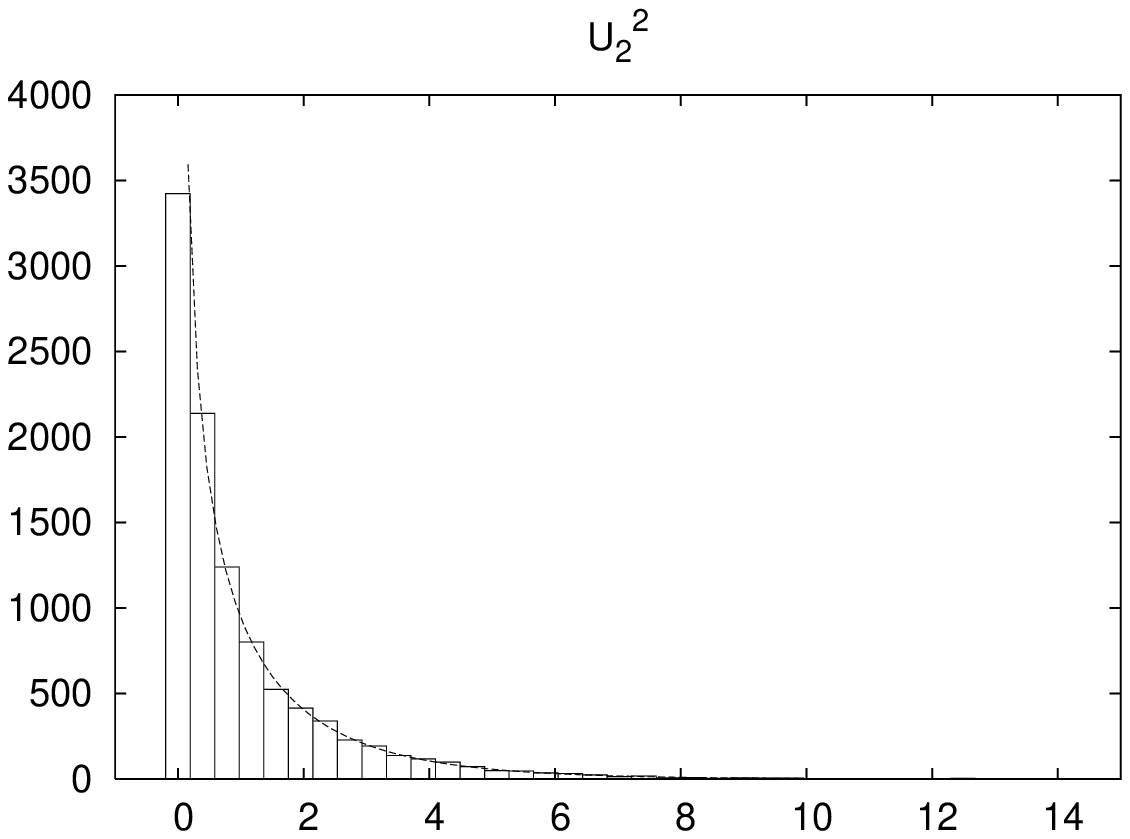}
\includegraphics[width=8cm]{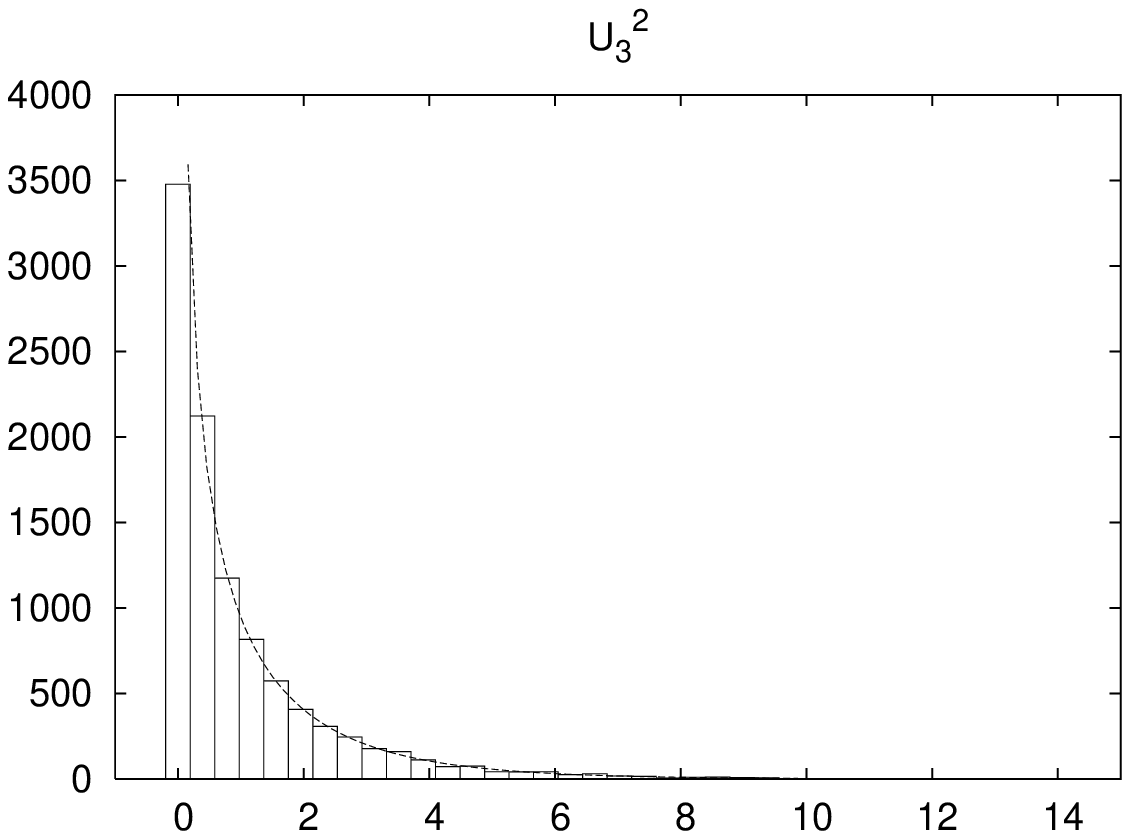}
\includegraphics[width=8cm]{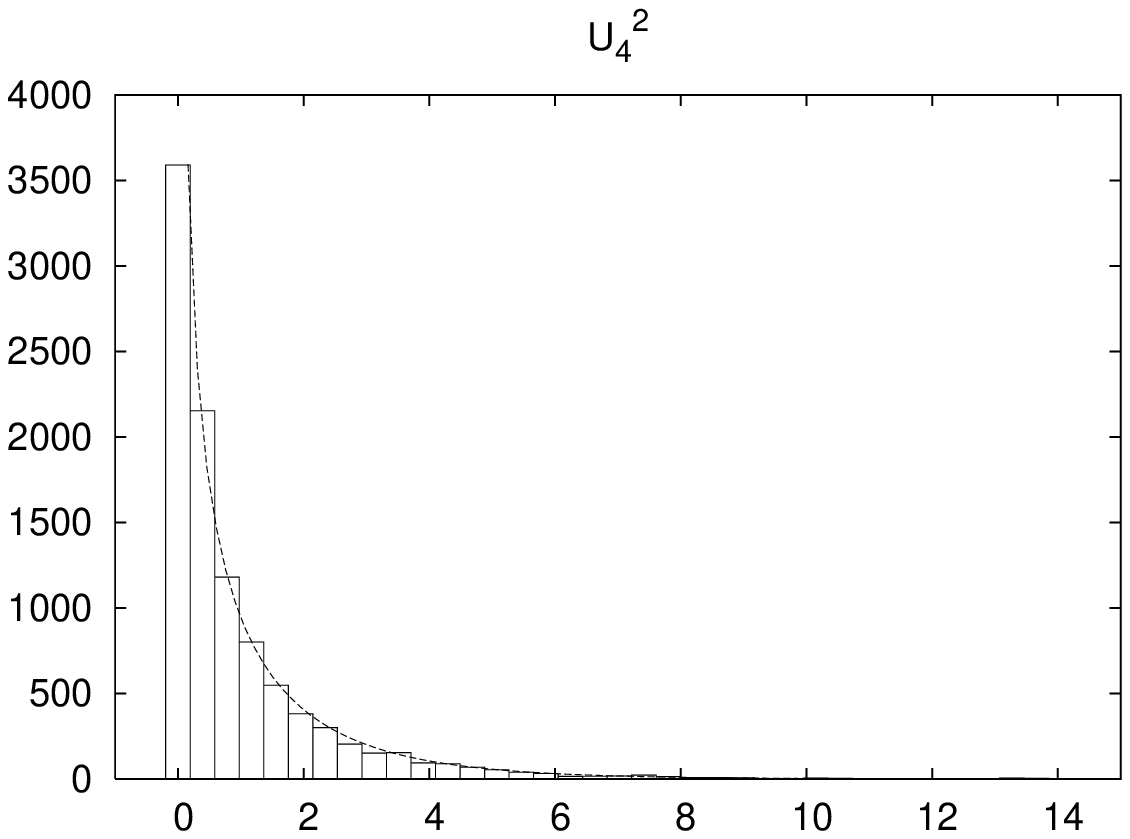}
\caption{Distributions of the $U_i^2$ statistics, from left to right, top to 
bottom $i=1,2,3,4$. They are obtained from 10000 simulated observations of a 
single-field observation with the VSA, consisting of a Gaussian CMB signal 
plus a noise level corresponding to $25\times4$~hr integration. The 
visibilities are binned in cells of 9$\lambda$, and the number of 
visibilities in each simulation is equal to 895. The dashed line is a 
$\chi_1^2$ function normalized to the total number of simulations.}
\label{fig2}
\end{center}
\end{figure*}

Given the simulated visibilities we decorrelate their real and 
imaginary parts as  explained in subsection \ref{sec:002}. After that, we 
calculate the distributions of the $U_i^2$ quantities (subsection 
\ref{sec:001}). As we have Gaussian simulations, if the amount of data is 
relatively large, these distributions must be very close to $\chi_1^2$ 
functions as it is explained above. The form of the distributions is shown in 
Fig. \ref{fig2} and they  are compared with $\chi_1^2$ functions normalized to 
the number of  simulations used  (dashed line). We have used 10000 
simulations of a VSA field, with a noise level 
corresponding to $25 \times 4$~hr for each one. This is the typical noise
level achieved in a single-field VSA observation \citep{grainge03}. 
We bin these data using a cell size of 9$\lambda$, to obtain 895 binned 
visibilities. In Table \ref{tab1} the mean value and 
the standard deviation ($\sigma$) of the distributions is compared 
with the same quantities of the $\chi_1^2$ distribution. 

As mentioned above, the mean value of the $U_i^2$ must be equal to 1 
independently of the number of data. From Table \ref{tab1} we see that the 
convergence to this value is very accurate. The value of the standard 
deviation must be $\sqrt{2}$ only when the number of data tends to 
infinity (asymptotic case). In the case we are simulating 
(895 visibilities) we see that the values of the standard deviation 
are very close to the $\chi_1^2$ values (only $U_4^2$ 
seems to separate slightly from the asymptotic value). The VSA fields we are
analysing have a number of visibilities between $\sim$1000 and $\sim$4000 
and then the approximation to the $\chi_1^2$ functions will be better 
than the case shown in Table \ref{tab1}. Because of that, we can 
approximate the $U_i^2$ to $\chi_1^2$ distributions in a very
accurate way.  We then suppose the distributions to be $\chi_1^2$ 
for the $U_i^2$ statistics or $\chi_k^2$ for the $S_k$. We have 
applied to these simulations the method described for the phases 
and we also find $\chi_1^2$ distributions for the $U_i^2$ statistics. 
The advantage of this assumption is that we do not have to make 
simulations to calculate the statistical distributions for every VSA 
field for the Gaussian or uniform null hypothesis. We have checked that the
distribution of the $U_i^2$ statistics is also very close to $\chi_1^2$ 
functions even when the number of data is $\sim$200. In the following 
sections we cut data with low signal-to-noise eigenvalues and the 
statistics are calculated with $\sim$200 data; also in this case, 
the distributions of the statistics will then be approximated by 
$\chi^2$ functions.

\begin{table}
\begin{center}
\caption{Values of the mean ($\langle U_i^2 \rangle$) and the standard 
deviation ($\sigma$) of the statistics $U_i^2$ for 10000 Gaussian CMB plus 
noise simulations of our template experiment. They are compared with the 
corresponding asymptotic values (for $\chi_1^2$), displayed in last column.}
\begin{tabular}{c c c c c c }
\hline
          & $U_1^2$ & $U_2^2$ & $U_3^2$ & $U_4^2$ & $\chi^2_1$  \cr
\hline
$\langle U_i^2 \rangle$             
           & 0.9863  & 0.9983  &  1.0370 & 1.0131  &   1.0000   \cr

$\sigma$   & 1.4077  & 1.4426  &  1.4848 & 1.5265  &   1.4142   \cr
\hline
\end{tabular}
\label{tab1}
\end{center}
\end{table}

\section{Non-Gaussian simulations}
\label{section_4}

In this section, given our template of a VSA observation, we 
generate non-Gaussian simulations of the visibilities and add the 
(Gaussian) VSA noise to them. 
We analyse three kinds of non-Gaussian simulations. First, we 
consider simulations created by means of the Edgeworth 
expansion in the visibility space; in second place, 
we analyse a cosmic strings simulation; finally,
$\chi^2$ simulations are considered as an exercise to estimate
the power of the test to constrain the $f_{NL}$ parameter. 

In order to characterize the power of the method in each of
these cases, we use the 
``power of the test''. Roughly speaking, the power, $p$, of a given test at a 
certain significance level $\alpha$ ($0<\alpha<1$) is parameterized as the 
area of the alternative distribution function outside the region 
which contains the $1-\alpha$ area of distribution function of the null 
hypothesis (the Gaussian case). Thus, a large value of $p$ for a prefixed 
small value of $\alpha$ indicates that the two distributions have a small 
overlap, so we can distinguish between them (e.g. \citealt{barreiro}). For 
definiteness, in what follows we adopt for the significance levels 
the values $\alpha = 0.05/0.01$ (i.e. significance levels of 5\% and 1\% 
respectively), so we shall quote these pair of values for each one 
of the cases considered. 

\subsection{Edgeworth Expansion}

We first construct simulations with a certain degree of non-Gaussianity by 
using the Edgeworth expansion \citep{martinez}. We then analyse these 
simulations and calculate the power of the test to discriminate between a 
Gaussian distribution (the null hypothesis) and a distribution with skewness 
and kurtosis injected via the Edgeworth expansion (the alternative hypothesis).
The aim is to quantify which signal-to-noise level is required to detect 
a certain degree of non-Gaussianity in the data with our method.

We have two options in preparing these simulations:
we can inject the skewness and kurtosis either in  real space or in visibility
space. 
If we include the non-Gaussianity through the Edgeworth expansion in real
space, this non-Gaussianity is 
diluted when we transform to the Fourier (or visibility) space 
(this is illustrated in Appendix A). 
Thus, we decide to inject the skewness/kurtosis
directly in Fourier space, and calibrate our method in terms of the ability to
detect deviations of the moments of the signal-to-noise eigenmodes of 
the visibilities with respect to the Gaussian case. A similar approach has 
been adopted by other authors (e.g. in \citealt{rocha} they explore the 
non-Gaussianity of the visibilities simultaneously to the estimation of 
the power spectrum, by using a modified non-Gaussian likelihood function).

To use non-Gaussian simulations with given skewness/kurtosis in Fourier
space generated from the Edgeworth expansion is not motivated by any 
physical model, but it is a reasonable approach to calibrate the power of a 
non-Gaussianity test because these non-Gaussian maps are easy to prepare 
with specified statistical properties. Moreover, systematic effects could 
introduce a non-Gaussian signal in the visibilities which could 
be mimic with the Edgeworth expansion. The probability 
distribution of these systematic effects could be, for example, asymmetric 
and, in a first approximation, this fact can be modeled by a distribution with 
a non-zero skewness. 
In addition, the distribution could be more peaked or flat 
that the Gaussian one, and that can be modeled by a distribution with a 
non-zero kurtosis. In this manner the Edgeworth simulations are taken as a 
benchmark to quantify the power of our method to detect some kind of 
non-Gaussian signals.

These non-Gaussian simulations are generated as follows. We 
generate a realization of independent values $y_j$ with zero mean, unit 
variance, and skewness and kurtosis $S$ and $K$, respectively. For 
definiteness, we adopt here the values $S = K = 1$. As in subsection 
\ref{sec:gauss}, the real parts of the visibilities are 
$V_i^R = \sum_j L_{s,ij}^R y_j$. The imaginary parts are 
generated in an analogous way. After that, we add different 
noise levels according to a VSA observation. We then decorrelate 
the data of the simulations and calculate the power of the tests.

\begin{figure}
\begin{center}
\includegraphics[width=8cm]{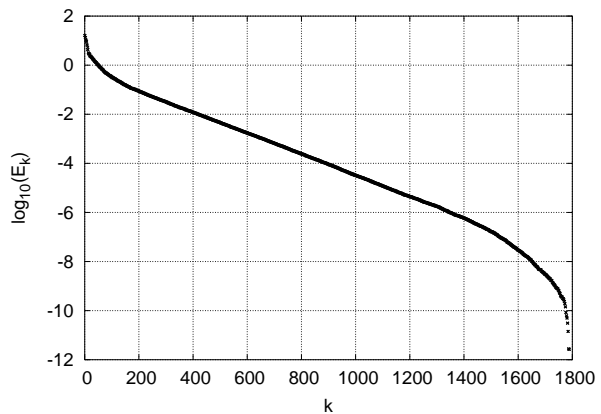}
\caption{Eigenvalues $E_k$ obtained from the template observation used in 
Figure \ref{fig2}. We plot the values associated with both the real and 
imaginary parts of the  visibilities (see the  text for details). We have 
$895\times2$ eigenvalues, only 420 of which have a value above 0.01, which 
corresponds to a signal-to-noise of $(0.01)^{1/2} = 0.1$.}
\label{fig3}
\end{center}
\end{figure}

Again, we use the same VSA template observation, binned into cells of side 
9$\lambda$ to study the power of our tests. Figure \ref{fig3} shows the 
logarithm of the eigenvalues associated with the real visibilities (those 
resulting from the diagonalization of $C^R_{s,ij}/(\sigma_i \sigma_j )$) and 
those associated with the imaginary parts (the diagonalization of 
$C^I_{s,ij}/(\sigma_i \sigma_j )$). Only the $23 \%$ of the data 
have an eigenvalue signal-to-noise higher than 0.01; that is, a
signal-to-noise ratio higher than 0.1.

\begin{figure}
\begin{center}
\includegraphics[width=8cm]{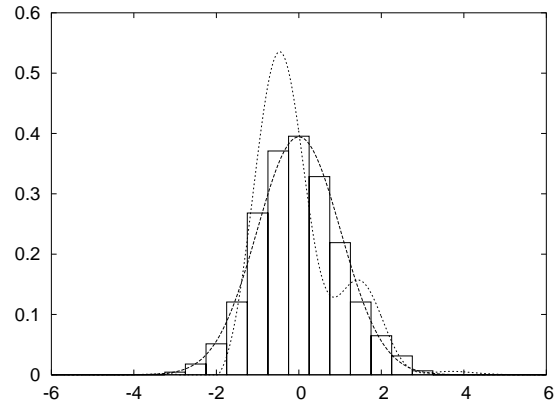}
\caption{Dotted line: distribution function created by the Edgeworth expansion 
with $S=1.0$ and $K=1.0$. Dashed line: Gaussian distribution. Histogram: 
distribution obtained from a single simulation (signal plus noise) involving 
all the eigenvalues, for a noise level corresponding to 
a $25 \times 25\times4$~hr integration. The three distributions have zero 
mean and unit variance. The non-Gaussianity in the single simulation is 
hardly seen in its histogram; however, it is easily detected when we used our 
method, as shown in Table \ref{tab3}.}
\label{ng_vs_g}
\end{center}
\end{figure}

First, we consider the case of a low signal-to-noise ratio, and we
adopt the noise levels corresponding to a single-field VSA observation
($25 \times 4$~hr, \citealt{grainge03}).
We analyse the non-Gaussian simulations with different cuts in the eigenvalues;
that is, we  include in the analysis only those eigenmodes whose eigenvalue is 
higher than a value $E^{\rmn{cut}}$. In Table \ref{tab2} is shown the power 
of the $U_1^2$, $U_2^2$, $U_3^2$ and $U_4^2$ statistics to discriminate between
a Gaussian  and a non-Gaussian distribution. It can be demonstrated that the
simulations are constructed such that $y_i$ has zero mean and unit variance,
so only the $U_3^2$ and $U_4^2$ distributions are notably different from a
$\chi_1^2$ distribution (see section \ref{sec:001}). In fact, $U_2^2$ also 
deviates slightly from a $\chi_1^2$ distribution when the analysed data have
some degree of kurtosis ($U_2^2$ is related to $\hat{\mu}_2^2$, i.e.\ to the 
fourth power of the data), but, as we can see in Table \ref{tab2}, this 
deviation is very low. The corresponding probability function is shown in Fig. 
\ref{ng_vs_g} (dotted line) and  is compared with a Gaussian function 
(dashed line). We have used 5000 simulations in this computation. Note that 
although the input skewness and kurtosis is 1.0, the output simulations have a 
mean skewness equal to $0.96 \pm 0.07$ and a mean kurtosis equal to 
$0.85 \pm 0.32$; i.e.\ we are analysing some simulations with kurtosis 
$\sim 0.5$. So we expect less power in the detection of the kurtosis that 
in the detection of the skewness. The first column in  Table \ref{tab2} 
indicates the minimum value of the signal-to-noise eigenvalues that we use 
to perform the analysis.  We indicate the square root of the latter quantity 
(signal-to-noise ratio) in parentheses. The second column is the number of 
eigenmodes that we use to calculate the statistics; that is, the number of 
eigenvalues such that $E_k > E^{\rmn{cut}}$. Finally, the last columns indicate
the power of the different statistics (in percentages) when the significance 
level is $5 \% / 1 \%$. We see that our test is not able to detect  
non-Gaussianity when every eigenmode is included in the analysis. When 
eigenmodes with low signal-to-noise ratio are excluded the power is improved 
but the values are very poor. This tells us that the noise level of a 
single VSA pointing ($25 \times 4$~hr) is too high to detect the non-Gaussian 
simulations we have constructed via the Edgeworth expansion.

\begin{table*}
\begin{minipage}{10cm}
\begin{center}
\caption{Power of $U_1^2$, $U_2^2$, $U_3^2$ and $U_4^2$ to discriminate between
a Gaussian  and a non-Gaussian distribution obtained from the 
Edgeworth expansion with input skewness and kurtosis equal to 1.0 (see the text
for a discussion about the actual skewness and kurtosis of the simulations). 
The template observation corresponds to an integration time of $25\times4$~hr.
The first column shows the value used for $E^{\rmn{cut}}$, and the second
column displays the number of remaining visibilities above the previous
threshold.
For each statistic is shown the power (as a percentage) when the significance
level is 5\%/1\%. }
\begin{tabular}{c r c c c c}
\hline
$E^{\rmn{cut}}$($\sqrt{E^{\rmn{cut}}}$)& Num.&$U_1^2$&$U_2^2$&$U_3^2$&$U_4^2$\cr
\hline             
      0.0 (0.0)      & 1790& 5.10/1.12 & 4.92/1.06& 9.06/2.46 & 6.44/ 1.90\cr
      0.1 (0.32)     &  188& 5.12/0.82 & 5.96/1.54&26.20/13.26&12.30/ 7.48\cr
      0.2 (0.45)     &  133& 5.08/0.92 & 6.68/1.70&31.26/17.68&13.76/ 8.76 \cr
      0.3 (0.55)     &  107& 4.84/1.04 & 6.38/1.74&32.06/18.46&13.70/ 9.70 \cr
      0.4 (0.63)     &   89& 5.06/0.94 & 6.56/2.14&33.64/20.22&15.02/10.90 \cr
      0.5 (0.71)     &   76& 5.82/1.10 & 7.14/2.30&34.22/20.56&14.82/10.94 \cr
\hline
\end{tabular}
\label{tab2}
\end{center}
\end{minipage}
\end{table*}

We therefore now explore  the case of an integration time of  
$25 \times 25\times4$~hr, which is comparable to the whole
signal-to-noise level achieved in the dataset presented in \cite{dickinson}.
In this case, the noise levels are reduced by a factor $5$ with respect to the
previous case. The results are shown in Table \ref{tab3}. As is to be expected,
the detection of non-Gaussianity is better. For example, we could detect our 
non-Gaussian simulations with a power equal to $\sim$99$\%$ (for the 
$U_3^2$ statistic)  with the signal-to-noise achieved during $2500$ hours 
and analysing eigenmodes with eigenvalues higher than 0.4 or 0.5. In this 
last case the distribution of all the data (every eigenmode) for one 
simulation is shown by the histogram in Figure \ref{ng_vs_g}. We see that this 
distribution is closer to a Gaussian  (dashed line) than the initial one 
(dotted line) for two reasons: first, the loss of skewness and kurtosis by the 
construction of the Edgeworth simulations (the actual skewness and kurtosis of 
the simulations is lower than the input values, see above) and, second, by the 
addition of Gaussian noise. When we only plot the autovalues higher than 0.4 or
0.5, the histogram is closer to the dotted line; that is, we see the 
non-Gaussianity.

\begin{table*}
\begin{minipage}{10cm}
\begin{center}
\caption{Power of $U_1^2$, $U_2^2$, $U_3^2$ and $U_4^2$ to discriminate between
a Gaussian  and a non-Gaussian distribution obtained from the Edgeworth 
expansion with input skewness and kurtosis equal to 1.0 (see the text for a 
discussion of the actual skewness and kurtosis of the simulations). The 
template observation corresponds to an integration time of 
$25\times25\times4$~hr. The columns have the same meaning as 
in Table~\ref{tab2}.}
\begin{tabular}{c r c c c c}
\hline
$E^{\rmn{cut}}$($\sqrt{E^{\rmn{cut}}}$)& Num.&$U_1^2$&$U_2^2$&  $U_3^2$  & $U_4^2$  \cr
\hline             
      0.0 (0.0)      & 1790& 5.38/0.86&5.52/1.22 &57.96/36.08&18.20/10.04  \cr
      0.1 (0.32)     &  513& 5.62/1.22&6.48/1.62 &95.92/88.62&35.58/25.64  \cr
      0.2 (0.45)     &  441& 5.94/1.52&7.36/1.82 &97.66/91.94&36.24/26.82  \cr
      0.3 (0.55)     &  402& 5.60/1.08&7.60/2.22 &98.26/93.28&37.42/28.52 \cr
      0.4 (0.63)     &  371& 5.90/1.54&7.52/2.02 &98.72/94.28&38.56/29.68 \cr
      0.5 (0.71)     &  245& 5.52/1.40&7.80/2.38 &98.82/94.44&39.24/30.80 \cr
\hline
\end{tabular}
\label{tab3}
\end{center}
\end{minipage}
\end{table*}

In the tables shown in this section we have  presented only the power of the
$U_i^2$ statistics. In the case we are studying, $S_3$ has less power than 
$U_3^2$ because $S_3=U_1^2+U_2^2+U_3^2$  also has information about $U_1^2$ and
$U_2^2$, which do not have power and  compensate for the power of $U_3^2$. 
However,  $S_3$  is better than the $U_3^2$ statistic if we also have 
detection of non-Gaussianity in $U_2^2$ and/or $U_1^2$. For the same reason, 
$S_4$ is slightly better than $U_4^2$ because it has combined information about
$U_4^2$ and $U_3^2$. In our case, however, $S_4$ is not better than $U_3^2$, so
we  show only the $U_i^2$ statistics.

Finally, we have analysed the Edgeworth simulations with the test applied to 
the phases (subsection \ref{sec:3_3}). We have found that the non-Gaussianity 
of this kind of simulation is not detected with the phases method; that is, 
the phases are compatible with a uniform distribution. A deeper analysis of the
phases shows that the moments $\hat{\mu}_{\alpha}$ (section \ref{sec:unif}) 
are sligthly different from the values corresponding to a uniform 
distribution. However, the quantity, $n$ of data with high signal-to-noise 
eigenvalues is not large enough to give values of the  $U_i^2$ statistics 
that differ sufficiently for the values obtained from a uniform distribution, 
so the power of the test has very low values. (The distribution for the 
phases can be calculated analytically by assuming, for example, a distribution 
for the real and imaginary parts given by equation (25) in \citealt{martinez}. 
We have found  this ditribution to oscillate about the uniform one, but the 
moments are very close to those of the uniform distribution.)

\subsection{Detectability of Cosmic Strings}

\begin{figure}
\begin{center}
\includegraphics[width=8.5cm]{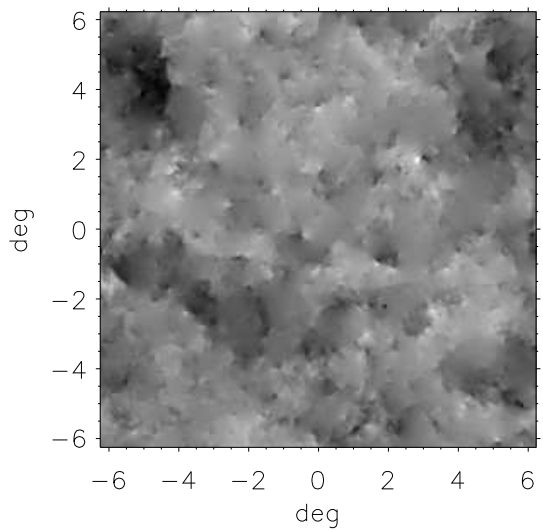}
\caption{String map analysed in this paper.}
\label{strings_map}
\end{center}
\end{figure}

In this section, we probe if our method is able to detect the non-Gaussian
signature introduced by cosmic strings. To this end,
we apply our method to the analysis of the string simulation
shown in Fig. \ref{strings_map} \citep{bouchet}. We first analyse the 
simulation without noise. In this way we can learn what happens when we have 
only the string map. After that we add different noise levels to see how the
method is able to detect the strings. 

Although in theory the correlation matrix depends only of the power spectrum
and the beam (see expression \ref{eq:002}), the (finite) pixel size of the real
string map we are analysing and the irregular coverage of the visibility plane
could introduce noticeable imprecisions. The correlation matrix of the strings 
is then calculated using Gaussian maps simulated with the same pixel size: 
we calculate the power spectrum of the string real map and construct Gaussian 
real maps with this power spectrum (we use 40000 Gaussian simulations). From 
these Gaussian maps we calculate the corresponding visibility values 
(expression \ref{eq:005}). The correlation of these visibility simulations is 
that used for the analysis of the strings. Note that in the expression 
(\ref{eq:002}) there is no hypothesis concerning the statistic of the 
temperature field. The only hypothesis is the homogeneity of the field. 
 
The template observation of the VSA is the same as in previous sections (895
binned visibilities in 9$\lambda$ cells).

\subsubsection{Noiseless Map of Strings}

In this case we diagonalize the correlation matrix of the signal because we do
not have noise. With this diagonalization we decorrelate the data in a similar 
way as  is done in  sections \ref{sec:002} and \ref{sec:004}.

\begin{figure}
\begin{center}
\includegraphics[width=8.5cm]{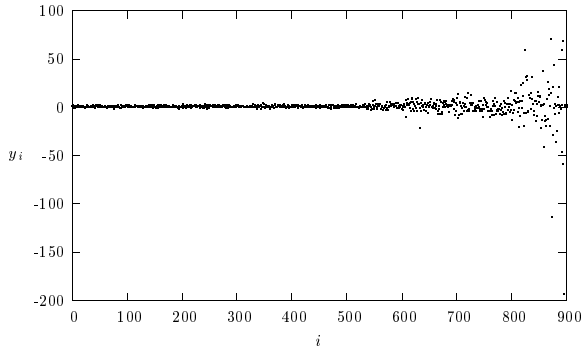}
\caption{Values of $y_i$  associated with the real part of the visibilities 
for an observation of the string map presented in Fig.~\ref{strings_map} 
using our VSA template without adding noise.}
\label{fig_y_s}
\end{center}
\end{figure}

We analyse the real and imaginary parts of the visibilities obtained from the 
string map without noise. The values found for the statistics are: 
$U_1^2 \simeq 1.4 \cdot 10^3$, $U_2^2 \simeq 3.4 \cdot 10^8$, 
$U_3^2 \simeq 2.9 \cdot 10^{13}$ and $U_4^2 \simeq 4.4 \cdot 10^{18}$. The 
estimated moments, $\hat{\mu}_\alpha$, are $\hat{\mu}_1 \simeq -0.89$, 
$\hat{\mu}_2 \simeq 6.2 \cdot 10^2$, $\hat{\mu}_3 \simeq  -3.1 \cdot 10^5$ and 
$\hat{\mu}_4 \simeq 2.5 \cdot 10^8$. These values are clearly incompatible with
a Gaussian realization. The $y_i$ quantities (see expression \ref{eq:y_k}) 
associated with the real parts of the string visibilities are shown in the 
Fig. \ref{fig_y_s} (those associated with the imaginary parts have similar 
features). Every $y_i$ has an associated signal eigenvalue $E_i$ whose 
value decreases with $i$. We see, then, that for the strings the absolute value
(or \emph{amplitude}) of $y_i$ grows when $E_i$ decreases (this explains why 
the moments $\hat{\mu}_\alpha$ are so large). This feature shows that the
$y_i$ are not well decorrelated or normalized.

\begin{figure}
\begin{center}
\includegraphics[width=8.5cm]{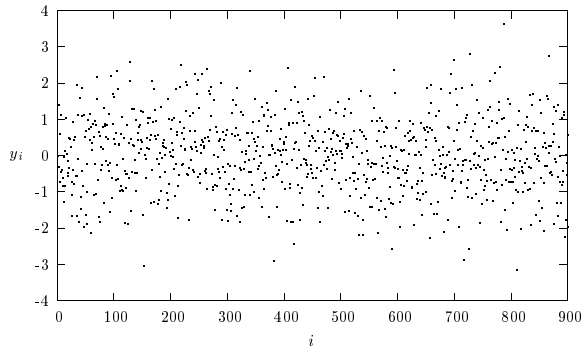}
\caption{Same as in Fig.~\ref{fig_y_s}, but for a simulated Gaussian CMB map.}
\label{y_s_gauss}
\end{center}
\end{figure}

We analyse a Gaussian map with the same  procedure. Figure
\ref{y_s_gauss} shows the resulting $y_i$ quantities for this Gaussian case. We
see that they have a width of order of unity. 
Thus, we conclude that the behaviour of Fig. \ref{fig_y_s} is a feature of the 
non-Gaussianity of the string signal. The fact that the $y_i$ quantities are 
not well decorrelated or normalized indicates that we are not able to 
estimate their correlation matrix accurately. 
This seems to be because we have a model with a very large cosmic variance, and
the correlation matrix (or equivalently the power spectrum) cannot be well 
estimated from only one realization. Moreover, we are estimating the power
spectrum with the expression that maximizes the likelihood under the 
hypothesis of Gaussianity for the temperature field \citep{bond2000}. But
the temperature field is not Gaussian for the strings, and so the 
estimated power spectrum for the strings (that which maximizes the 
likelihood in this case) can  differ from the one we are 
using. For a study of the dependence 
of the power spectrum on non-Gaussianity see \cite{amendola}.

In the Gaussian case, however, we are able to calculate the power spectrum 
and then we decorrelate and normalize  the visibilities properly. In this way, 
the $y_i$ quantities have a width equal to unity, i.e.\ 
$\langle y_i^2 \rangle = 1$, as   shown in  Figure \ref{y_s_gauss}. 
Summarizing, when there is no noise, in principle we should 
detect the strings by means of all the $U_i^2$ statistics (even $U_1^2$ and 
$U_2^2$) because their values are not compatible with a $\chi_1^2$ 
distribution (see Table \ref{tab1}).

Finally, we have applied the smooth tests of goodness-of-fit developed by 
Neyman to the phases of the visibilities of the noiseless map of strings
(section \ref{sec:3_3}). The $U_i^2$ values are: $U_1^2=3.64$ ($94.36\%$), 
$U_2^2=4.59$ ($96.78\%$), $U_3^2=0.32$ ($42.84\%$) and $U_4^2=0.13$ 
($28.16\%$). And for the $S_i$ statistics: $S_1=3.64$ ($94.36\%$), $S_2=8.23$ 
($98.37\%$), $S_3=8.55$ ($96.41\%$) and $S_4=8.68$ ($93.04\%$). The probability
of obtaining a lower or equal statistic value  under the null hypothesis (the 
visibilities are Gaussian) is showed in parentheses. 

In the following subsection we discuss the effect of Gaussian noise on
the detectability of strings with our method.

\subsubsection{String map with different noise levels}

The features in Fig. \ref{fig_y_s} change when (Gaussian) noise is added. The 
behaviour for low values of $i$ remains unchanged because these values are 
associated with high signal-to-noise values so that these data are dominated 
by the signal. However, the behaviour for high $i$ is dominated by the noise 
(low signal-to-noise eigenvalues) so that the behaviour will resemble 
that of a Gaussian signal. 

The analysis with different noise levels is given in Table \ref{tab4} (noise
corresponding to an integration time equal to $25\times4$~hr) and Table 
\ref{tab5} (noise corresponding to an integration time equal to 
$25\times25\times4$~hr). We have analysed 5000 simulations involving the 
string simulation plus a noise realization. The analysis is done for different 
cuts of the eigenvalues ($E^{\rmn{cut}}$). The power to distinguish between a 
Gaussian distribution 
and the strings plus noise when the significance level is $5\%/1\%$ is shown
in the tables.

\begin{table}
\begin{center}
\caption{Power of the $U_i^2$ ($i=1,2,3,4$) statistics for simulated
  observations of the strings map using a noise level 
corresponding to an integration time of $25\times4$~hr. 
The columns indicate the power in percentage when the 
significance level is $5\%/1\%$.}
\begin{tabular}{l c c c c}
\hline
$E^{\rmn{cut}}$&  $U_1^2$  &   $U_2^2$  &   $U_3^2$  & $U_4^2$    \cr
\hline             
      0.       & 5.16/0.82 &  5.72/1.20 &  4.92/1.00 & 4.50/0.98  \cr
      0.1      & 3.72/0.58 &  6.00/0.86 &  2.28/0.36 & 1.70/0.28  \cr
      0.2      & 3.10/0.30 &  4.16/0.44 &  1.70/0.24 & 1.20/0.34  \cr
      0.3      & 5.10/0.66 &  3.36/0.22 &  1.48/0.26 & 1.24/0.30  \cr
      0.4      & 1.46/0.14 &  1.78/0.12 &  1.08/0.36 & 1.28/0.30  \cr
      0.5      & 1.00/0.10 &  2.04/0.08 &  0.80/0.18 & 0.90/0.22  \cr
\hline
\end{tabular}
\label{tab4}
\end{center}
\end{table}

\begin{table}
\begin{center}
\caption{Power of the $U_i^2$ ($i=1,2,3,4$) statistics for simulated
observations of the string map using a noise level corresponding 
to an integration time of $25 \times 25\times4$~hr. The columns indicate the 
power in percentage when the significance level is $5\%/1\%$.}
\begin{tabular}{l c c c c}
\hline
$E^{\rmn{cut}}$&  $U_1^2$  &    $U_2^2$ &   $U_3^2$  & $U_4^2$    \cr
\hline             
      0.       &  5.30/0.98 &  24.26/8.44& 4.20/0.86& 4.28/1.04 \cr
      0.1      &  6.56/0.84 & 79.56/49.32& 1.58/0.12& 1.30/0.22 \cr
      0.2      &  7.18/0.92 & 86.72/59.60& 1.08/0.08& 1.24/0.34 \cr
      0.3      &  2.94/0.26 & 87.62/60.44& 0.80/0.04& 0.70/0.10 \cr
      0.4      &  3.74/0.10 & 90.74/64.16& 0.48/0.02& 0.66/0.08 \cr
      0.5      &  3.20/0.12 & 92.90/69.58& 0.48/0.08& 0.38/0.08 \cr
\hline
\end{tabular}
\label{tab5}
\end{center}
\end{table}

The results in Table \ref{tab4} indicate that the VSA noise of a 
single field observed during $25\times4$~hrs is too high for detecting the 
strings with our method. When the integration time is multiplied by 25 
(Table \ref{tab5}) we start to detect the strings by means of the $U^2_2$ 
statistic. It is important to note 
that, even if the data were well decorrelated, the $U^2_2$ statistic would also
be a indicator of non-Gaussianity. For example, given realizations of 
normalized and independent data with kurtosis $K \neq 0$, the mean value of 
$U^2_2$ is equal to $(K+2)/2 \neq 1$. Another example of the use of $U^2_2$ to
detect non-Gaussianity can be found in \citet{aliaga3b} in the formalism of the
multinormal analysis.    

As a test, we repeat the same procedure using a Gaussian simulation instead
of the string map and, as expected, we obtain a negligible power.

%
%

\subsection{$\chi^2$ simulations} \label{sec:chi2}

Nowadays, non-Gaussian simulations constructed by the addition of a Gaussian
field and its square have acquired a notable relevance because they could 
represent perturbations produced in several inflationary 
scenarios (for a review on the subject, see \citealt{bartolo} and
references therein). The parameter which measures the coupling with
the non-linear part is denoted by $f_{NL}$ and, then, an analysis to
study the power of our method to detect this parameter seems to be
very appropriate. Realistic simulations in the flat-sky approximation
would require to develop appropriate software (see
e.g. \citealt{liguori} for simulations on the full sphere), and this
is beyond the scope of the present work. 
However, we can estimate the power of our method in detecting
a non-zero $f_{NL}$ component using a simple approximation.
We follow \cite{aghanim03}, and we generate what they call ``$\chi^2$ maps''
by assuming that the observed temperature contrast field on sky 
($\Delta (\bmath{x}) \equiv \Delta T(\bmath{x})/T$) can be
written as

\[
\Delta (\bmath{x}) = \Delta_L(\bmath{x}) + 
f_{NL} (\Delta_L^2(\bmath{x}) - \langle \Delta_L^2(\bmath{x}) \rangle)
\]
where $\Delta_L$ is the Gaussian (linear) component. 
In reality, this approximation is only valid in the Sachs-Wolfe regime  
(see e.g. \citealt{komatsu2001}), but
we use it as a toy model in our exercise of the study of the 
detection of the $f_{NL}$ parameter\footnote{
Note that in the Sachs-Wolfe
regime, $\Delta (\bmath{x})= \Phi (\bmath{x})/ 3$, so the 
definition of the non-linear coupling parameter in our $\chi^2$ model
gives $\Phi (\bmath{x}) = \Phi_L(\bmath{x}) + 
(f_{NL}/3) (\Phi_L^2(\bmath{x}) - \langle \Phi_L^2(\bmath{x})
\rangle)$. Therefore, the coupling parameter used by
\cite{komatsu2001} and \cite{cayon_a} is related with our definition
as $f_{NL}/3$, and the one used by \cite{smith} is $2f_{NL}/9$.
}.

\begin{figure}
\begin{center}
\includegraphics[width=8cm]{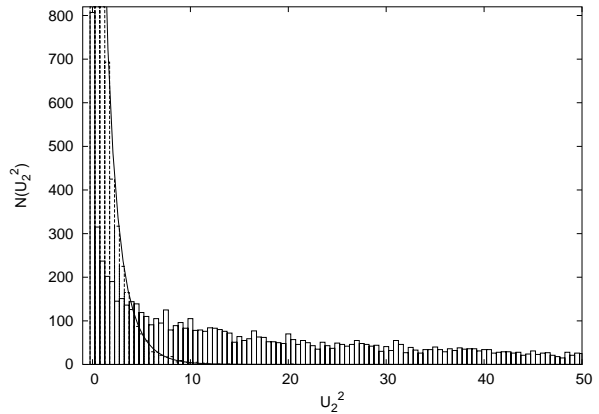}
\caption{Distributions of $U_2^2$ for the Gaussian case (dashed boxes) and for 
the non-Gaussian case with $f_{NL}=10000$ (solid boxes). The solid line is a 
$\chi^2_1$ function normalized to the number of simulations (10000
simulations). (For the sake of clarity, the histogram for the Gaussian
case has been cut at the top.)}
\label{fig_chi2}
\end{center}
\end{figure}

We first consider the ideal case where there is no instrumental noise,  
and we have analyzed different sets of simulations with $f_{NL}$ values 
$0$ (Gaussian case), $100$, $1000$ and $10000$. We have run $10000$ 
simulations for every case. Only the case with $f_{NL}=10000$ has been 
detected via the $U_2^2$ statistic. 
Fig.~\ref{fig_chi2} shows the distribution function of the $U_2^2$ 
statistic for the Gaussian case (dashed boxes), compared with the corresponding
one for the non-Gaussian case with $f_{NL}=10000$ (solid boxes). From
here, we infer that the power of this statistic to discriminate  
between the non-Gaussian simulations and the Gaussian ones is about the 
$79\%$ ($72\%$) with a significance level of $5\%$ ($1\%$). 
Analysing the same previous simulations but adding the noise level 
corresponding to an observation time of $25 \times 25 \times 4$~hr, we
find that the power of our statistic is reduced roughly by a
factor 2.

\cite{smith} constrained the $f_{NL}$ parameter 
by using the bispectrum of the VSA extended array data (note that this 
is exactly the same configuration that we have adopted to build our visibility 
template).  With their definition of $f_{NL}$, they found an upper
limit of 7000 with a confidence level of $95\%$. In the Sachs-Wolfe
regime, this would correspond to a value of 31500 according to our $f_{NL}$
definition. However, this value can not be direclty compared with
the one obtained in our work, because our model does not 
correspond to the realistic case which was considered in \cite{smith}.

\section{Discussion and Conclusions}\label{section_5}

In this paper we have presented a method of searching for non-Gaussianity in 
data from interferometric  CMB observations, directly in the visibility space. 
This method can be adapted to other interferometric experiments 
(e.g.\ CBI and DASI) if we know the correlation matrices of the signal and 
of  instrumental noise. Note that in this paper, we have dealt with 
decorrelated noise (which is usually the case for these experiments), but 
the case of correlated noise can be studied in an analogous way.

We have applied the method to work with the real and imaginary parts of 
the visibilities. The method tests whether they are Gaussian distributed. 
However, we have applied the method to the phases of the visibilities 
to test whether they are uniformly distributed, but we found that it is not
very sensitive for detecting the kind of non-Gaussianity we have analysed here.

We have integrated the signal-to-noise formulation into the smooth 
goodness-of-fit tests. In this way we can deal only with the data dominated 
by the signal we want to analyse. In the text it is noted that 
the correlation matrices of the signal-to-noise eigenmodes and of the signal 
are descomposed as the product of a matrix and its transpose (subsection 
\ref{sec:002} and section \ref{sec:gauss}). This decomposition is analogous 
to that of Cholesky. This latter decomposition is computationally faster than 
the one used here; however, the decomposition we use allows us to deal with
better quality data, that is, with a higher signal-to-noise ratio. The 
analysis with the Cholesky decomposition takes every eigenvalue; 
that is, $E^{\rmn{cut}}=0$.

It is important to stress that  smooth goodness-of-fit tests do not 
require the data to be on a regular grid, but can be applied to any data set. 
In that sense the test is perfectly adapted to interferometric data because 
the coverage of the $UV$ plane is neither regular nor complete. The method 
could therefore also be applied by selecting those visibilities in a certain 
$UV$ range, thus allowing  study of the non-Gaussianity as a function of  
angular scale. 
 In addition, although the method has been presented here as a tool to 
study the non-Gaussian properties of the sky signal, it could also be used as 
a powerful diagnostic for detecting systematics in the data, as we pointed out 
in Section 6.1. For short integration periods, a stack of visibilities will be 
dominated by instrumental noise, so this method could be used to trace the 
presence of spurious signals in the data (e.g. those comming from cross-talk 
between antennas), or to study the correlation properties of the noise.

Summarizing, to study the power of our method in detecting non-Gaussian 
signals on the sky, we have analysed three kinds of simulations.
First we have analysed non-Gaussian visibilities created by 
inserting some degree of skewness and kurtosis ($S=K=1$) with the Edgeworth 
expansion directly in Fourier space. 
Using the VSA as a reference experiment, we have shown the performance of
the method in detecting those levels of non-Gaussianity in the data with
realistic values of the integration time for this experiment. 
All these results can be easily adapted to other instruments, just by
rescaling the integration times according to the square of the ratio of the 
different sensitivities. 
In addition, we have also shown the performance of the
test in the detection of the non-Gaussian signal introduced by cosmic strings. 
Even though those kind of signals are usually detected using real-space 
statistics or wavelets \citep{hobson99,barreiro}, we have demonstrated that 
the signal-to-noise eigenmodes approach allows us to detect them because 
the method is very sensitive to the characterization of the covariance matrix. 
In this particular case, due to the non-Gaussian nature of the
strings, a complete decorrelation can not be achieved and the statistics
show huge deviations with respect to the Gaussian case. 
For the case of the VSA, this translates into the fact that
cosmic strings can  hardly be detected in single-field observations
(integration times of $25\times4$~hr), but they could be detected 
(if present) using this method, given the current sensitivity achieved by the
whole data set published by the VSA team (with a sensitivity of 
$\sim$25 $\times 25\times4$~hr integration). 
Finally, we have studied the power of the method to detect a 
non-zero $f_{NL}$ component in a toy model based on $\chi^2$ maps.


\section*{Acknowledgements}

We would like to thank F. R. Bouchet for kindly providing the string map, and
to R. Rebolo and the anonymous referee for useful comments. We also
acknowledge Terry Mahoney for revising the English of the
manuscript. JAR-M acknowledges the hospitality of the IFCA during two
visits. AMA acknowledges the IAC for its hospitality during a
visit. RBB acknowledges the MCyT and the UC for a Ram\'on y Cajal
contract. We acknowledge partial financial support from the Spanish MCyT  
project ESP2002-04141-C03-01.

\appendix

\section[]{Bispectrum of the Edgeworth expansion}

Suppose a square real map of pixel coordinates $\bmath{x}$ and pixel data 
$y_{\bmath{x}}$. These values are generated via the Edgeworth expansion
such that they are independent, normalized and have skewness $S$. 
The map has $N_{\rmn{side}}$ pixels on a side and the area of the pixel is
$\Delta^2$. We now perform its Fourier transform:
\[
\tilde{a}_{\bmath{k}} = \Delta^2 \sum_{\bmath{x}} y_{\bmath{x}} e^{i\bmath{k}\bmath{x}}
\]
whose bispectrum is
\[
\langle
\tilde{a}_{\bmath{k_1}} \tilde{a}_{\bmath{k_2}} \tilde{a}_{\bmath{k_3}}
\rangle =
\Delta^6 \sum_{\bmath{x_1} \bmath{x_2} \bmath{x_3}}
\langle y_{\bmath{x_1}} y_{\bmath{x}_2} y_{\bmath{x}_3} \rangle
e^{i\bmath{k}_1\bmath{x}_1}
e^{i\bmath{k}_2\bmath{x}_2}
e^{i\bmath{k}_3\bmath{x}_3}
\]

Taking into account that $y_{\bmath{x}}$ are independent and 
$\langle y_{\bmath{x}}^3 \rangle = S$ we have
\[
\langle
\tilde{a}_{\bmath{k_1}} \tilde{a}_{\bmath{k_2}} \tilde{a}_{\bmath{k_3}}
\rangle =
N_{\rmn{side}}^2 \; \Delta^6 \; S \; 
\delta_{\bmath{k}_1+\bmath{k}_2+\bmath{k}_3,\bmath{0}}
\]
where $\delta_{\bmath{k}_1,\bmath{k}_2}= N_{\rmn{side}}^{-2}
\sum_{\bmath{x}} e^{i(\bmath{k}_1- \bmath{k}_2) \bmath{x}}$ is the Kronecker 
delta. The Fourier modes could be normalized to the desired spectrum:
$a_{\bmath{k}} = 
C_k^{1/2} \tilde{a}_{\bmath{k}}/(N_{\rmn{side}}\Delta^2)$ such that 
$\langle a_{\bmath{k}} a_{\bmath{k}}^* \rangle 
= C_k \delta_{\bmath{k}_1,\bmath{k}_2}$. Finally, the (normalized) bispectrum
is given by

\[
\frac{\langle a_{ \bmath{k}_1} a_{ \bmath{k}_2} a_{\bmath{k}_3} \rangle}
{\langle |a_{\bmath{k}_1}|^2 \rangle^{1/2}
\langle |a_{\bmath{k}_2}|^2 \rangle^{1/2}
\langle |a_{\bmath{k}_3}|^2 \rangle^{1/2}} = 
\frac{S}{N_{\rmn{side}}}\delta_{\bmath{k}_1+\bmath{k}_2+\bmath{k}_3,\bmath{0}}.
\]
Note that the Gaussianity associated to the third moment 
in the Fourier space is 
reduced by a factor $N_{\rmn{side}}$ compared to that in real space.

\section[]{Skewness and kurtosis of a given eigenmode from 
the Edgeworth expansion}

When we introduce some degree of non-Gaussianity (skewness or kurtosis) 
by means of the Edgeworth expansion in the visibilities, it is clear that
each eigenmode will show a different degree of non-Gaussianity which
will depend on the associated signal-to-noise eigenvalue. This 
relation can be obtained as follows. 

As it is explained in the main text, we 
generate a realization of independent values $y_j$ with zero mean, unit 
variance, and skewness and kurtosis $S$ and $K$, respectively. 
As in subsection \ref{sec:gauss}, the real parts of the visibilities are 
$V_i^R = \sum_j L_{s,ij}^R y_j$. The imaginary parts are 
generated in an analogous way. After that, we add the corresponding different 
noise levels and decorrelate the data by using the transformation given by
equation (\ref{eq:003}). Then, bearing in mind the definition of the $E_k$ 
quantities, it can be easily demonstrated that for the final decorrelated and
normalized data we have 

\[
\langle y_k^3 \rangle = \bigg ( \frac{E_k}{E_k+1} \bigg )^{3/2} S \quad ; \quad
\langle y_k^4 \rangle - 3 = \bigg ( \frac{E_k}{E_k+1} \bigg )^{2} K.
\]

As one would expect, the skewness and the kurtosis are lost if the (Gaussian) 
noise dominates the data ($E_k \ll 1$). In contrast, the skewness and the 
kurtosis are preserved for the case  $E_k \gg 1$. In the previous expressions, 
the angled brackets ($\langle \cdot \rangle$) denote an average over 
realizations.

\label{lastpage}

\end{document}